%% file: archive2.tex
\documentclass[12pt]{article}

\usepackage{amsmath,amssymb}
\usepackage{feynmp}
\usepackage{graphicx}

\def\be{\begin{equation}}
\def\ee{\end{equation}}
\def\bea{\begin{eqnarray}}
\def\eea{\end{eqnarray}}

\def\dk{\int\!\!\frac{d^4k}{(2\pi)^4i}}

\def\dk4{\int\!\!\frac{{\rm d}^4k}{(2\pi)^4i}}	%
\makeatletter
    
    \@addtoreset{equation}{section}
\makeatother

\begin{document}
\setlength{\baselineskip}{18pt}
\begin{titlepage}


\vspace*{1cm}

\begin{center}
{\LARGE\bf The Higgs Particle and Higher-Dimensional Theories}\footnote{to be published in Progress of Theoretical and Experimental Physics} 
\end{center}
\vspace{25mm}

\begin{center}
{\large
C.S. Lim 
\footnote{e-mail : lim@lab.twcu.ac.jp}
}
\end{center}
\vspace{1cm}
\centerline{{\it Department of Mathematics, Tokyo Woman's Christian University, 
Tokyo 167-8585, Japan}}

%
%
\vspace{2cm}
\centerline{\large\bf Abstract}
\vspace{0.5cm}

In spite of the great success of LHC experiments, we do not know whether the discovered ``standard model-like" Higgs particle is really what the standard model predicts or a particle some new physics has in its low energy effective theory. Also the long-standing problems concerning the property of Higgs and its interactions are still there, and we still do not have any conclusive argument of the origin of the Higgs itself. 

In this article we focus on higher-dimensional theories as new physics. First we give a brief review of their representative scenarios and closely related 4-dimensional scenarios. Among them, we mainly discuss two interesting possibilities of the origin of the Higgs: Higgs as a gauge boson and Higgs as a (pseudo) Nambu-Goldstone boson. 

Next, we argue that theories of new physics are divided into two categories, i.e. theories with normal Higgs interactions and those with anomalous Higgs interactions. Interestingly both of two candidates concerning the origin of the Higgs mentioned above predict characteristic ``anomalous" Higgs interactions, such as the deviation of the Yukawa couplings from the standard model predictions. Such deviations can be hopefully investigated by the precision tests of Higgs interactions at the planned ILC experiment. 

Also discussed is the main decay mode of the Higgs, $H \to \gamma \gamma$. Again, theories belonging different categories are known to predict remarkably different new physics contributions to this important process.       

\end{titlepage} 

\newpage

\section{Introduction}

\subsection{What the discovery of the standard model-like Higgs particle means} 

We have heard of the very impressive results by LHC experiments that a new particle withe the mass 126 (GeV), which behaves as the Higgs particle of the standard model (SM), was discovered at CERN. So far the property of the particle seems to be consistent with the Higgs of the SM, but it will be premature to conclude whether it is really the Higgs of the SM or a standard model-like Higgs particle predicted by some theory of new physics (physics beyond the standard model). In this article we discuss what the discovery of that particle means in the context of higher-dimensional theories with extra dimensions. 

\subsection{New physics and its ``decoupling limit"} 

Though we are interested in the possibility that the standard model-like Higgs is some particle predicted by new physics, it is at the same time a clear fact that the SM is very successful theory at low energies ($E \leq M_{W}$). We thus expect that any theory of new physics reduces into the SM at low energies. To be more precise, we can think of \\ 
``decoupling limit" \\ 
of the new physics theories. The limit is achieved by sending the typical mass (energy) scale of new physics, say $M$, to infinity: $M \to\infty$. In that limit, various new particles predicted to exist in the new physics with masses of ${\cal O}(M)$ are expected to ``decouple" from the low energy sector \cite{AC}, and the theories are expected to reduce into the SM. From field theoretical point of view, the effects of new particles at low energies are described by operators with mass dimension higher than 4 (in the viewpoint of the 4-dimensional (4D) space-time), ``irrelevant operators" whose Wilson coefficients are suppressed by the inverse powers of $M$ ($1/M^{2}$ in the case of operators with mass dimension 6). Thus the effects of new heavy particles are suppressed and they ``decouple" from the low energy effective lagrangian. 

Let us note that according to the argument based on the dimensional analysis, marginal or relevant operators with mass dimensions equal or less than 4 may be affected by the presence of heavy new particles. The argument of \cite{AC} is that such effects only affect the renormalization procedure and can be absorbed into the bare parameters in the process of renormalization. Let us note, however, the hierarchy problem, to be more precise the problem of quadratically divergent quantum correction to the Higgs mass, is exactly concerning the relation between the observed Higgs mass and corresponding bare mass(-squared) parameter. Thus when we consider the hierarchy problem the contributions of new heavy particles should be seriously taken into account. This is why super-partners in supersymmetry (SUSY) theory can play important role in the solution to the hierarchy problem. Also in a type of the higher-dimensional gauge theory, ``gauge-Higgs unification (GHU)" \cite{Manton}, \cite{Fairlie}, \cite{Hosotani} discussed later, the summation over the contributions of all Kaluza-Klein (KK) modes makes the quantum correction to the Higgs mass-squared finite, thus solving the hierarchy problem \cite{HIL}. 

For instance, in the MSSM (Minimal Supersymmetric Standard Model) two Higgs doubles $H_{u}, H_{d}$ are introduced to preserve SUSY. The lighter CP-even neutral scalar particle denoted as a linear combination, $\sqrt{2}[\cos \alpha (\mbox{Re}H^{0}_{u} - \frac{v_{u}}{\sqrt{2}}) - \sin \alpha (\mbox{Re}H^{0}_{d} - \frac{v_{d}}{\sqrt{2}})]$, is identified with the SM-like Higgs. The angle $\alpha$ is fixed by the parameters in the Higgs potential and $v_{u,d}$ are the VEV's of two Higgs doublets. The couplings of the SM-like Higgs with matter fermions and gauge bosons generally contain parameters $\alpha$ and $\beta$, where $\beta$ represents the ratio of the VEV's, $\tan \beta \equiv v_{u}/v_{d}$. The decoupling limit of MSSM is achieved by sending the SUSY breaking mass scale $M_{SUSY}$ to infinity. In this limit it turns out that just a linear combination of two Higgs doublet which alone develops the VEV $v = \sqrt{v_{u}^{2} + v_{d}^{2}}$ becomes the lighter scalar and is identified with the SM-like Higgs, as we naturally expect. Correspondingly, in this limit $\alpha \to \beta - \frac{\pi}{2}$ and all interactions are known to reduce to those we expect in the SM, as is anticipated from the decoupling theorem \cite{AC}. We therefore conclude that the decoupling limit of the MSSM is just the SM, although in MSSM, in contrast to the case of the SM, the Higgs mass is ``calculable", since the quartic coupling of the Higgs potential gets contribution only from the ``D-term" and the Higgs mass at the tree level is more or less equivalent to the weak gauge boson masses. 

Now let's turn to the case of higher dimensional theories. Because of the presence of the extra dimension, we now get an infinite tower of KK modes with higher extra space momenta and therefore higher 4D masses. The non-zero KK modes are new heavy particles. Hence in higher dimensional theories the decoupling limit is achieved by 
\be 
\label{1.1} 
M_{c} = \frac{1}{R} \ \ \to \ \ \infty, 
\ee 
where $R$ is a generic size of the extra dimension, being the radius when the extra space is just a circle $S^{1}$, and 
$M_{c}$ is the corresponding ``compactification" mass scale. In this limit, all (non-zero) KK modes, having masses of the order of 
$M_{c}$, are expected to decouple from low energy effective lagrangian. 

As a new feature of higher dimensional gauge theory, it is basically possible to construct a theory in which Higgs-like particle does not exist in the decoupling limit, i.e. higgsless model \cite{Chaba}. In this type of theory the KK zero-mode with vanishing extra space 
momentum is excluded by imposing Dirichlet-type boundary condition for the Higgs field along an interval, as the 1-dimensional extra space. Thus all KK modes of the Higgs field have masses of the ${\cal O}(M_{c})$, while the violation of the unitarity of the scattering amplitude of the (longitudinal components of) weak gauge bosons is remedied at higher energies by the presence of a tower of massive no-zero KK modes of the gauge bosons. The decoupling limit of the higgsless theory does not contain SM-like Higgs. 

Now that we know that at least a SM-like Higgs has been discovered. The higgsless scenario thus seems to be excluded unless a theory with lower $M_{c}$ is possible without contradicting with various experimental data.
 
Also ruled out is the theory where, even though the decoupling limit is the SM, the modification of the Higgs sector (Higgs interaction) is so huge that it already contradicts with the reality. The GHU model 
with H-parity \cite{HOOS} probably is such sort of theory. The H-parity is a discrete symmetry under which among SM fields only Higgs field has an odd parity while other fields have even parities. This symmetry exists only when the weak scale and the compactification scale are comparable: 
\be 
\label{1.2}
x \equiv \frac{g_{4}}{2}\pi R v = \pi \frac{M_{W}}{M_{c}} = \frac{\pi}{2},  
\ee
where $g_{4}$ is the 4D gauge coupling constant. Note that even though $M_{c}$ is near to the weak scale, the masses of non-zero KK modes can be sufficiently large due to the presence of the warp factor when the theory is formulated on the Randall-Sundrum space-time \cite{Agashe}, \cite{Hosotani2}. Just as the R-parity implies in the MSSM, the H-parity implies that a single Higgs particle cannot decay into ordinary SM particles, thus making the Higgs stable. As the recently discovered SM-like Higgs particle decays into 2 photons, e.g., this interesting scenario seems to be inconsistent with the data. 

There still remain other interesting higher-dimensional scenarios. It should be noticed that some of these scenarios, not only being consistent with the data having the SM at its decoupling limit, but also predicts some characteristic deviations from the SM predictions, as we will see below.  

Even though the properties of the SM-like Higgs particle seem to be consistent with that of the SM Higgs, we still have a strong motivation to investigate further the properties of the Higgs. Namely, in the SM, in clear contrast to the sector of gauge interaction the Higgs sector is mysterious having several long-standing problems, such as the hierarchy problem, already mentioned and the problem of many arbitrary parameters in the sector of Higgs interactions, such as Yukawa couplings. We still have not understood the origin of the generation-dependent hierarchical fermion masses and 
generation (flavor) mixings. These problems all come from the fact that, in contrast to the case of the gauge interactions there is no guiding principle (symmetry ?) to restrict the interactions of the Higgs. Let us note that even though many of the new physics theories have been proposed mainly in order to solve the hierarchy problem, not all of them have mechanisms to restrict the Yukawa coupling. For instance, in MSSM there is no principle to restrict the couplings (except for gauge invariance) and the Yukawa couplings are free parameters to start with. 

These problems suggest that we have not (fully) understood the origin of Higgs itself. From such points of view, it will be of crucial importance to study how new physics theories predict on the Higgs interactions and to perform the precision tests of those interactions. As was mentioned above, some of higher dimensional theories and also some closely related 4D theories make very specific predictions concerning the Higgs interactions, i.e. ``anomalous Higgs interactions".     

Such precision tests will not be easy to perform at the LHC experiment and are expected to be done at planned ILC experiment.

\section{Brief review of higher dimensional theories} 

We now summarize very briefly some of the representative scenarios of new physics based on higher-dimensional theories. Also discussed are 4D scenarios closely related to higher dimensional gauge theory, i.e. the scenarios in which Higgs is a (pseudo) Nambu-Goldstone (NG) boson due to some global symmetries. Since solving the hierarchy problem has been main motive of many of new physics theories, we categorize the scenarios depending on their attitudes concerning the hierarchy problem. 

\subsection{Theory without the solution to the hierarchy problem}    

The representative and popular scenario of this type is so-called ``universal extra dimension (UED)". 
It is the straightforward extension of the SM to its higher dimensional version \cite{UED}. All SM particles are allowed to propagate 
universally in the higher-dimensional bulk space. Extra dimension is assumed to be an orbifold, torus divided by its discrete symmetry, such as $S^{1}/Z_{2}$ or $T^{2}/Z_{4}$, in order to realize a chiral (left-right asymmetric) theory needed to incorporate the SM. 
The theory has a discrete symmetry called KK-parity, under which particles of even(odd) KK numbers have even(odd) KK parities. Thus the lightest first KK mode, having odd KK parity, cannot decay into ordinary SM particles with even KK-parities and thus becomes stable, just as the LSP in MSSM with R-parity, and has a right to be the candidate of the dark matter. This KK-parity makes the production of a single first KK exited state impossible at ILC. On the other hand the production of second KK exited state with even KK-parity is possible through the KK-number violating processes induced by the quantum loop effects \cite{Cheng}. 

\subsection{Theories with the mechanism to solve the hierarchy problem not invoking any symmetry} 

We now discuss the scenarios formulated in higher-dimensional space-time aimed to solve the hierarchy problem, namely to explain naturally the hierarchy between the Planck scale $M_{pl}$ and $M_{W}$. The scenarios we discuss are the ADD model with large extra dimension \cite{ADD} and the Randall-Sundrum model with warped extra dimension \cite{RS}. In both scenarios, the SM particles live in a 4D space-time (a brane), not extra dimension, and only gravity propagates in the bulk. This property leads to the new types of solutions to the hierarchy problem, not seen in the theories in ordinary 4D space-time. The mechanism to solve the hierarchy problem does not invoke any symmetry. 

Let us first discuss the ADD model. When A. Einstein proposed unified theory of gravity and electromagnetism along the idea of Kaluza and Klein, the extra dimension was assumed to be small, i.e. $M_{c} \simeq M_{pl}$: ``small extra dimension". The basic reason is that in the unified field theory, electromagnetic interaction originates from gravity interaction. Thus the electric charge $e$ is inevitably proportional to the (square root of) Newton constant $G_{N}$. On the other hand the source of the gravity is the energy-momentum and the electric charge should be also proportional to the extra-dimensional momentum of the order of $M_{c}$. In this way, we get a relation 
\be 
\label{2.1}
e = \frac{4 \sqrt{\pi G_{N}}}{R} \ \ \to \ \ R = \frac{4 \sqrt{\pi G_{N}}}{e} \simeq 4 \times 10^{-32} \ (\mbox{cm}).  
\ee

In the ADD model, all the SM particles including the photon is assumed to live in the brane, and the size of the extra dimension is free from the constraint coming from the value of electric charge. In this scenario, the original higher dimensional Planck scale $M^{(0)}_{pl}$ is of the order of weak scale $M^{(0)}_{pl} \sim M_{W}$ and originally there is no hierarchy between the Planck scale and the weak scale. Such too small Planck scale may lead to unacceptably strong gravitational force. This potential difficulty is solved by assuming a large extra dimension. Namely, the ``gravitational flux" is spread toward the direction of the large extra dimension, thus recovering ordinary very weak 4D gravity. Assuming the presence of $n$-dimensional extra dimension, 4D Planck scale $M_{pl}$ is given as 
\be 
\label{2.2}
M_{pl}^{2} \sim M_{pl}^{(0)2+n}\cdot R^{n}.   
\ee
Assuming $M_{pl}^{(0)} \sim 1$ (TeV), we get $R \sim 0.1$ (mm) for $n = 2$ (for $n = 1$ obtained $R$ is unacceptably large). This is the scenario of ``large extra dimension". 

While the gravity at the distances larger than the size of compactification reduces to ordinary Newtonian gravity, at shorter distances, the gravity recovers the original higher dimensional one whose mass scale is not $M_{pl}$ but $M^{(0)}_{pl} \sim M_{W}$. Thus the most stringent bound for the ADD model comes from the date from LHC. A search for the anomalous jet + missing $E_{T}$ events at CMS has put a bound $M^{(0)}_{pl} > 3.0-5.0$ (TeV) for $n = 2-6$ (for higher $n$ the bound is less stringent) \cite{CMS}.  

This scenario, however, faces its own new hierarchy problem: the size of the extra dimension 0.1 (mm) for $n = 2$ means $M_{c} = 2 \times 10^{-2}$ (eV), which implies a new hierarchy $M_{c}/M_{W} \sim 10^{-13}$ !

Randall and Sundrum pointed out that such problem of new hierarchy in the ADD model can be solved once we allow that the bulk space-time is a curved one \cite{RS}. They assume that 5D space-time is anti-de Sitter with a negative cosmological constant while the extra space is compactified on an orbifold $S^{1}/Z_{2}$. The solution to the Einstein equation has a ``warp factor": 
\be 
\label{2.3} 
g_{MN} = 
\begin{pmatrix}
e^{- 2\kappa |y|}\cdot \eta_{\mu \nu} & 0 \\  
0 & 1  
\end{pmatrix}, 
\ee
where in the warp factor $e^{- 2\kappa |y|}$, $\kappa$ is a parameter of the theory of ${\cal O}(M_{pl})$ and $y$ is the extra space coordinate. The warp factor may be understood as the factor of ``space-like inflation". In fact it mimics the factor $e^{Ht} \ (H: \mbox{Hubble constant})$ in the inflationary universe caused by a positive cosmological constant. The sign difference of the cosmological constant may be attributed to the sign difference of the time and space components of the metric tensor. The absolute value $|y|$ in the warp factor is due to the orbifolding of the extra space $S^{1}/Z_{2}$. 
  
Assuming that the SM particles all live in the 4D brane at $y = \pi R$ (``visible brane"), the hierarchy between the Planck scale and the weak scale is naturally realized by the warp facitor without any hierarchy between the input parameters of the theory: 
\be 
\label{2.4} 
\frac{M_{W}}{M_{pl}} \sim e^{- \kappa \pi r_{c}}
\ee 
where $r_{c}$ is the radius of the circle. 

The theory predicts the presence of the non-zero KK modes of graviton, whose masses are spaced by ${\cal O}$(TeV) and their gravity couplings are suppressed not by $M_{pl}$ but by ${\cal O}$(TeV). Therefore, such new KK gravitons can be searched for at LHC as the resonance states in the $l^{+}l^{-}$ and $\gamma \gamma$ final states. The current lower bound on the KK graviton mass ranges from 0.9 (TeV) to 2.1 (TeV) depending on the parameter of the theory \cite{CMS2}, \cite{ATLAS}. 

By the way, how can this Randall-Sundrum model solve the hierarchy problem at the quantum level, i.e. the problem of quadratic divergence in the quantum correction to the Higgs mass, without relying on some symmetry ? Though an explicit argument cannot be found in the literature, the quadratic divergence will also be accompanied by the warp factor after the renormalization of the Higgs field as $\Lambda^{2} e^{- 2\kappa \pi r_{c}}$ and the divergence will be harmless.         
  
\subsection{Theories with the mechanism to solve the hierarchy problem invoking some symmetry} 

The hierarchy problem is concerning some very small number. The smallness of physical observable can be ``naturally" preserved under its quantum correction provided that some symmetry is enhanced in the action of the theory when that observable is switched off. This is because in the case the condition is met, even if the observable is induced at quantum level, it should be inevitably proportional to that small number, such that the correction goes away at the limit of exact symmetry. 

We now discuss a few representative scenarios of this sort in the context of higher dimensional theories. Some of them actually can be formulated in 4D space-time, but are closely related to a kind of higher dimensional gauge theory. 

\subsubsection{Gauge-Higgs unification} 

What we first discuss is the scenario of ``gauge-Higgs unification (GHU)".  Einstein attempted to unify at that time known gravity and electromagnetic interactions mediated by bosons with spin $s = 2, \ 1$ in the framework of 5D gravity theory. Now we know there is also Higgs interaction. So it is natural to expect that gauge and Higgs interactions mediated by $s = 1, \ 0$ bosons are unified in the framework of higher dimensional gauge theories. For instance in the simplest 5D U(1) gauge theory the gauge field $A_{M}$ is decomposed into 
\be 
\label{2.5} 
A_{M} = (A_{\mu}, \ A_{y})
\ee 
where $A_{\mu}$ corresponds to 4D gauge field (and its non-zero KK partners), while the extra space component $A_{y}$, to be more precise its KK zero mode, behaving as 4D scalar, is identified with the (SM-like) Higgs field. This is the scenario of GHU, whose idea is not new \cite{Manton}, \cite{Fairlie} and \cite{Hosotani}. In particular, Hosotani proposed a mechanism of dynamical spontaneous gauge symmetry breaking due to the VEV of the $A_{y}$ for non-Abelian case, ``Hosotani mechanism" \cite{Hosotani}.

The nice thing of this scenario is that it provides a new avenue for the solution to the hierarchy problem by virtue of higher dimensional local gauge symmetry under which $A_{y}$ transforms inhomogeneously: $A_{y} \to A_{y} + \partial_{y} \lambda$, $\lambda$ being $y$-dependent gauge parameter \cite{HIL}. We know photon never gets mass even at the quantum level, since local mass-squared operator $m_{A}^{2} A_{\mu}A^{\mu}$ is forbidden by local gauge symmetry. In the same way, the Higgs (the zero-mode of $A_{y}$) never has local mass-squared operator, thus eliminating the quadratic divergence. What was stressed in \cite{HIL} was the importance of the summation over all KK modes at the quantum correction of the Higgs mass. It is frequently argued that when we consider low energy effective theories only KK zero-modes should be taken into account, probably relying on the wisdom of decoupling theorem \cite{AC}. However, as has been already pointed out in the introduction, when we consider the hierarchy problem, heavy new particles play 
important roles. Also, we should note that a momentum cutoff spoils local gauge invariance. If we truncate the KK modes at some level, it is equivalent to the cutoff of the extra-dimensional momentum. This is why the summation over all KK modes is crucial to get the finite quantum correction to the Higgs mass.  

Actually the Higgs acquires a finite mass at the quantum level. This is because the zero-mode of $A_{y}$, i.e. the Higgs, has a physical interpretation as a Aharonov-Bohm (AB) phase or the phase of Wilson loop. Let us note that the VEV of Higgs 
in this scenario is nothing but a constant gauge field, which gives vanishing field strength and therefore seems to be just a pure gauge configuration. However, in the case in which the extra dimension is a non-simply-connected space like a circle $S^{1}$, the constant gauge field can be interpreted as a component of vector potential generated by the magnetic flux, penetrating inside the circle. Thus the Higgs field is not a pure gauge but has a physical meaning as the AB phase or the phase of the Wilson-loop $W$. At the quantum level, the Higgs potential is induced as (the real part of) the polynomial of $W$. 
Note that $W = \mbox{P}(e^{i\frac{g}{2} \oint A_{y} dy})$ is of course gauge invariant but global (not local) operator, which has nothing to do with UV divergence. This is why we get finite but non-vanishing Higgs mass in GHU, which disappears at ``decompactification limit $R \to \infty$", where the Wilson loop is trivial. The situation is very similar to the case of finite temperature field theory. At finite temperature the Coulomb potential of photon is known to have a mass of ${\cal O}(T) \ (T: \mbox{temperature})$. The mass disappears as $T \to 0$, corresponding to the decompactification limit.  

Another nice thing of GHU is that it may shed some light on the problem of arbitrary Yukawa couplings, for instance. 
In this scenario the Higgs is originally a gauge field. Therefore Yukawa coupling is gauge coupling to start with and if we succeed to construct a realistic model, the Yukawa couplings are expected to be constrained by the gauge principle. On the other hand, it is non-trivial question how the hierarchical fermion masses are realized starting from the gauge coupling, which is universal for all generations. Fortunately, as a new feature of higher-dimensional gauge theories, when orbifold $S^{1}/Z_{2}$ is adopted as the extra dimension so-called $Z_{2}$-odd bulk mass term of the form 
\be 
\label{2.6} 
- \epsilon (y) M \bar{\psi}\psi  
\ee
is allowed. Here $\epsilon (y)$ is sign function ($\pm 1$ depending on the sign of $y$), which mimics kink-like configuration of some scalar field and causes the localization of 
Weyl fermions at two different fixed points of orbifold depending on its chirality: the mode function of KK zero-mode of right and left-handed fermions behave as $\propto e^{-M|y|}$, \ $\propto e^{-M|y-\pi R|}$ ($R$ is the radius of $S^{1}$). Since the Yukawa coupling is the overlap integral of these mode functions of different chiralities, we eventually get the exponentially suppressed fermion masses behaving as ($M_{i}$ being flavor-dependent bulk masses)
\be 
\label{2.7} 
\sim M_{W} \ (\pi R M_{i})e^{- \pi RM_{i}} \ (R: \mbox{the radius of} \ S^{1})
\ee
for lighter (1st and 2nd) generations. We thus have a mechanism to realize the hierarchical fermion masses without any hierarchical structure of the parameters $M_{i}$. It is interesting to note that if we plot the logarithm of observed quark masses as 
the function of generation number they align along a straight line, roughly speaking. If we take this seriously, this suggests that quark masses were all equal before the exponential suppression, which is exactly what GHU implies starting from the universal Yukawa couplings. A study along this line of argumentation is now going on.     

For the scenario of GHU to be viable, we need to construct a minimal model based on this scenario, just as the MSSM in the case of SUSY. In this attempt, one non-trivial thing arises. In the case of MSSM, the SM was just made supersymmetric with the same gauge group 
SU(2) $\times$ U(1) for electroweak sector. In the case of GHU, however, this gauge group should be inevitably extended. This is because in this scenario Higgs is originally gauge field and therefore belongs to an adjoint repr. of gauge group, while as is well known the Higgs in the SM is SU(2) doublet, i.e. a fundamental repr. of SU(2). The breakthrough of this problem is to extend the gauge group a little. The simplest choice is to adopt SU(3) as the electro-weak sector. A minimal SU(3) GHU (electro-weak) model has been discussed \cite{KLY}.  

The minimal model is formulated in 5D space-time with an orbifold $S^{1}/Z_{2}$ as its extra dimension. The orbifold is defined by the identification of two points on the circle connected by $Z_{2}$ transformation, 
\be 
\label{2.8}
Z_{2}: \ \ y \ \ \to \ \ -y,   
\ee
where $y$ is the coordinate along $S^{1}$. We have two fixed points $y = 0, \ \pi R$, which are invariant under the transformation. Main motive to adopt the orbifold, not just a manifold like $S^{1}$, is by the ``orbifolding" a chiral theory is realized. By the identification under the $Z_{2}$ 
the degree of freedom of the extra space points becomes one half and correspondingly either right or left Weyl fermion survives as the fermion zero mode. 

Another merit to adopt the orbifold is by assigning different $Z_{2}$-parity for each element of irreducible repr. of SU(3), 
the gauge symmetry can be explicitly broken by the orbifolding \cite{Kawamura}, as is explained now.  

To realize the mechanism, we assign the $Z_{2}$-parities for the elements of SU(3) triplet repr. as follows 
\be 
\label{2.9} 
\Psi(- y)
= -{\cal P} \gamma^5 \Psi(y) \ \ \ 
({\cal P}= 
\begin{pmatrix} 
1 & 0 & 0 \\ 
0 & 1 & 0 \\ 
0 & 0 & -1 
\end{pmatrix} 
),  
\ee 
where the triplet $\psi$ contains quark fields 
\be 
\label{2.10}
\psi 
= 
\begin{pmatrix} 
u_{L} \\ 
d_{L} \\ 
d_{R}
\end{pmatrix}. 
\ee
The matrix ${\cal P}$ represents the $Z_{2}$-parities of the elements of the triplet. Let us note that as is seen in (\ref{2.9}), $Z_{2}$ has a aspect of chiral transformation for fermions. Thus the zero-mode, having even $Z_{2}$-parity, of upper two elements of $\psi$ are left-handed fermion, while the zero-mode of the lowest element is right-handed. Thus a chiral theory needed to accommodate the SM is realized. Now it is clear that ordinary bulk mass term of the form $M \bar{\psi}\psi$ is not allowed since $Z_{2}$ contains the chiral transformation. This is why the $Z_{2}$-odd bulk mass term (\ref{2.6}) is introduced to be consistent with the orbifolding.  

The KK zero-mode of the 4D gauge boson sector is given as 
\be 
\label{2.11}
A^{(0)}_{\mu}  
= \frac{1}{2} 
\begin{pmatrix}  
W^{3}_{\mu} + \frac{B_{\mu}}{\sqrt{3}} & \sqrt{2} W^{+}_{\mu} & 0 \\ 
\sqrt{2}W^{-}_{\mu} & - W^{3}_{\mu} +  \frac{B_{\mu}}{\sqrt{3}} & 0 \\  
0 & 0 & -\frac{2}{\sqrt{3}}B_{\mu} 
\end{pmatrix}.  
\ee
Namely, only the gauge bosons connecting the elements of $\psi$ with the same $Z_{2}$-parities have even $Z_{2}$ parity and therefore the zero-modes. It is now clear that the zero-mode sector of the gauge bosons is exactly what we need in the SM. In this way, SU(3) gauge symmetry is broken into SU(2) $\times$ U(1) by the orbifolding \cite{Kawamura}. 

The zero-mode sector of the 4D scalar is given as 
\be 
\label{2.12} 
A^{(0)}_{y}  
=  \frac{1}{\sqrt{2}} 
\begin{pmatrix}  
0 & 0 & \phi^{+} \\ 
0 & 0 & \phi^{0} \\ 
\phi^{-} & \phi^{0 \ast} & 0 
\end{pmatrix}.  
\ee
We find that this time only the part of ``broken generators" $G/H$ ($G =$ SU(3), \ $H$ = SU(2) $\times$ U(1)) has the zero-modes. 
This is because $A_{\mu}$ and $A_{y}$ should have opposite $Z_{2}$-parities just as $x^{\mu}$ and $y$ have. The ``off-diagonal" elements of (\ref{2.12}) just correspond to the SU(2) doublet of Higgs field in the SM. Thus by orbifolding we just get both of necessary gauge fields and Higgs doublet of the SM, and nothing else. It seems to suggest that the adjoint repr. of SU(3) has been prepared in order to accommodate the gauge-Higgs sector of the SM: $8 \to 3 + 1 + 2 \times 2$.  

\subsubsection{Higgs as a pseudo Nambu-Goldstone boson} 

As the possible symmetries in order to solve the hierarchy problem, so far we have discussed supersymmetry and gauge symmetry (Higgs as a gauge boson in the case of GHU). There remains the third possibility, i.e. global symmetry: Higgs as a (pseudo) Nambu-Goldstone (NG) boson. A representative scenario based on this idea is ``little Higgs (LH)".  

Quite interestingly, there seems to be close relation between GHU and LH scenarios, although the LH is a theory in the 4D space-time.  

There are a few ``circumstantial evidences" to imply such close relation: 

\noindent (a) In both scenarios, the gauge group of the SM is enlarged to some larger group G, which is broken to some subgroup H, and Higgs is identified with the NG boson of G/H in the case of LH, and $A_{y}$ of G/H in GHU as we have seen. 

\noindent (b) The coupling of $A_{y}$ to fermions in 5D GHU takes a form $g \bar{\psi} (i\gamma_5) \psi \cdot A_y$, 
which mimics the pseudo scalar coupling of NG bosons, such as pions. 

\noindent (c) Both have shift symmetries,
\be 
\label{2.13} 
A_y \to A_y + \partial_{y} \lambda \ \ (\mbox{for GHU}), \ \ G \to G + \mbox{const.} \ \ (\mbox{for LH}), 
\ee
where the former transformation is higher-dimensional local gauge transformation and the latter one is the transformation of some global symmetry.  

Actually the LH scenario was proposed being inspired by a scenario so-called ``dimensional deconstruction", which may be regarded as a `latticized" GHU, as we will see below. So, once we put the scenario of dimensional deconstruction as the bridge between GHU and LH, 
their mutual relation becomes more solid.     

One may wonder how global and local symmetries can be ``closely related". The point is that in the LH and also in the dimensional deconstruction there is a repetition of a global symmetry, SU(m) $\times$ SU(m) $\times \cdots \times$ SU(m) = (SU(m)$)^{N}$ with gauge parameters $\lambda_{i} \ \ (i = 1,2, \ldots N)$. On the other hand in the GHU, higher dimensional local gauge symmetry is described by a gauge parameter $\lambda (y)$. If we treat the integer $i$ of $\lambda_{i}$ as a ``discretized extra space coordinate" they can be identified. This argument suggests that dimensional deconstruction (and also LH) may be understood, roughly speaking, as a sort of ``latticized" GHU. 

\noindent \underline{Dimensional deconstruction} 

Before going into the LH scenario, we briefly discuss the scenario of dimensional deconstruction \cite{Dimensional deconstruction}, and its close relationship with the GHU. 

As the remarkable features of the scenario, \\ 
・Higgs is a pseudo NG boson, a bound state of fermions, just as pions in QCD.　\\ 
・There is a repetition of gauge symmetries: $(G \times G_{s})^{N}$ \ \ \ ($G =$ SU(m), \ $G_{s}$ = SU(n)). \\
・The quantum correction to the Higgs mass is finite (for $N \geq 3$) without relying on SUSY.      

The model has $N$ pairs of (, say left-handed,) Weyl fermions as matter fields with the bi-fundamental repr.s of     
\bea 
&&(m, \bar{n}), \ \ (\mbox{SU}_{i}(m), \mbox{SU}_{i}(n)) \nonumber \\ 
&&(\bar{m}, n) \ \ (\mbox{SU}_{i+1}(m), \mbox{SU}_{i}(n)) \ (i = 1 - N),  
\label{2.14} 
\eea 
where periodic boundary conditions SU$_{N+1}(m)$ = SU$_{1}(m)$ etc. are imposed. The structure of the theory is represented by 
so-called ``moose diagram" shown in Fig.1, where each oriented line corresponds to Weyl fermion with either $(m, \bar{n})$ or $(\bar{m}, n)$ repr., depending on the blob denoting the gauge group $G_{s}$ is at the right or left of the line.   

\begin{figure}[htb]
\begin{center}
\input{figure1.tex}\\ 
\end{center}
\end{figure}

Both of SU(m) and SU(n) are asymptotically free gauge symmetry with typical mass scales $\Lambda, \ \Lambda_{s}$. Assuming $\Lambda \ll \Lambda_{s}$, the SU(n) interaction becomes strong at higher energy, thus forming a bound states of these Weyl fermions just as the hadrons in QCD. At lower energies, $E \leq \Lambda_{s}$, the effective low energy theory is described by pseudo scalars a la pion $\pi_{i}$ as the (pseudo) NG bosons due to the spontaneous breaking of the chiral symmetry SU$_{i}(m) \times$ SU$_{i+1}(m)$:  
\be 
\label{2.15}  
U_{i} = e^{i\frac{\pi^{a}_{i}T_{a}}{f} } \ \ (T_{a}: \mbox{generators of SU(m)}),  
\ee 
where $f$ corresponds to the pion decay constant. The non-linear realization $U_{i}$ behaves as a bi-fundamental repr. of the chiral symmetry: i.e. $(m, \bar{m}) \ \mbox{under} \ (\mbox{SU}_{i}(m), \mbox{SU}_{i+1}(m))$.   
 
Actually SU(m) has been gauged and the effective action is a gauged non-linear sigma model: 
\be 
\label{2.16}
S = \int \ d^{4}x \ (- \frac{1}{2g^{2}}\sum_{j=1}^{N} tr (F_{j}^{\mu \nu})^{2} 
+ f^{2} \sum_{j=1}^{N} tr [(D_{\mu}U_{j})^{\dagger}(D^{\mu}U_{j})]),  
\ee
with the covariant derivative 
\be 
\label{2.17}
D_{\mu}U_{j} = \partial_{\mu}U_{j} -i A^{j}_{\mu}U_{j} + i U_{j}A^{j+1}_{\mu}.  
\ee 
Here $A^{j}_{\mu}$ are gauge bosons of SU(m) and $F_{j}^{\mu \nu}$ are their field strengths.    
     
This is nothing but 5D SU(m) pure non-Abelian gauge theory, with extra-space being latticized: $U_{j}$ may be regarded as the link variable (``Wilson line" along the extra dimension). Indeed, in the simplified Abelian case 
\be 
\label{2.18} 
D_{\mu} U_{j} \ \to \ \ i [\partial_{\mu}\pi - gf (A^{j+1}_{\mu} - A^{j}_{\mu})]U_{j},  
\ee 
and identifying the lattice spacing $a$ of the extra dimension as 
\be 
\label{2.19}
a = \frac{1}{gf},  
\ee 
$D_{\mu} U_{j}$ just corresponds to the field strength $F_{\mu y}$ in the GHU: 
\be 
\label{2.20}
D^{\mu}U_{j} \ \ \to \ \ 
\partial_{\mu}A_{y} - \partial_{y} A_{\mu} = F_{\mu y} \ \ 
(\pi \equiv A_y). 
\ee
Thus, the extra dimension has been constructed by the dynamics in 4D space-time. It should be noticed that above $\Lambda_{s}$ the theory recovers original renormalizable theory, in clear contrast to the non-renormalizable higher dimensional gauge theory, which the GHU is formulated on. 

The theory has gauge symmetry $(\mbox{SU}(m))^{N}$ which is spontaneously broken into a single SU(m):
\be 
\label{2.21}
\langle U_{i} \rangle = 1 \ \ \to \ \ g_{i} \langle U_{i}\rangle g_{i+1}^{\dagger} = 
\langle U_{i} \rangle, \ \ \mbox{only when} \ g_{i} = g_{i+1},  
\ee 
where $g_{i}$ is an element of SU$_{i}(m)$. Thus among $U_{i} \ \ (i = 1,2, \ldots, N)$, $N-1$ pieces are ``eaten" by the Higgs mechanism, and there remains only one physical  (pseudo) NG boson, which is identified as the Higgs. The remaining Higgs 
\be 
\label{2.22}
Tr (U_{1} U_{2} \ldots U_{N}),
\ee
is invariant under the gauge transformation, $U_{1} U_{2} \ldots U_{N} \ \ \to \ \ g_{1}(U_{1} U_{2} \ldots U_{N}) g_{1}^{\dagger} \ \ (g_{N+1} = g_{1})$, and therefore cannot be ``gauged away" by taking unitary gauge. 　 
  
In Abelian case, $U_{1} U_{2} \ldots U_{N} = e^{\frac{i}{f}(\pi_{1} + \ldots \pi_{N})} \ (\phi = \frac{\pi_{1} + \pi_{2} + \ldots \pi_{N}}{\sqrt{N}})$, and the field $\phi$ just corresponds to the zero-mode of $A_{y}$, or the phase of Wilson-loop, in GHU: $W = e^{i\frac{g}{2} \oint \ A_{y}dy}$. 

The field $\phi$ is pseudo NG boson. In fact, at the quantum level its potential is induced: 
\be 
\label{2.23} 
V(\phi) = - \frac{9}{4\pi^{2}}g^{4}f^{4} \sum_{1}^{\infty} 
\frac{\cos (\frac{2n\sqrt{N}\phi}{f})}{n(n^{2}N^{2}-1)(n^{2}N^{2}-4)} + \mbox{constant}. 
\ee
This potential is known to be finite for $N \geq 3$, which can be easily shown bu using relations such as $\sum_{n} \cos (\frac{2\pi}{N}n) = 0 \ (-\frac{N}{2} < n \leq \frac{N}{2})$. Let us note that such summation over all possible $n$ just corresponds 
to the summation over all KK modes in the GHU, which led to the finite Higgs mass \cite{HIL}. Thus we may also understand that the scenario of dimensional deconstruction provides very reasonable regularization scheme of the KK mode sum; We do not have to take sum over 
infinite number of KK modes but just a few modes are enough to guarantee the finiteness of the Higgs mass.  

The potential reduces to that in GHU \cite{Hosotani}, \cite{HIL}, \cite{KLY} in the limit $N \to \infty$: 
\be 
\label{2.24} 
V(A_y) = \frac{9}{4\pi^{2}} \frac{1}{(2\pi R)^{4}}
\sum_{n=1}^{\infty} \frac{\cos (ng A_{y} 2\pi R)}{n^{5}}. 
\ee 

\noindent \underline{Little Higgs} 

The purpose of the little Higgs (LH) scenario is to construct a 4D theory including SM, where Higgs is a pseudo NG boson, while the quadratic divergence of the quantum correction to the Higgs mass cancels out without relying on the SUSY. Though LH is a scenario inspired by the dimensional deconstruction, in this approach the NG boson needs not to be a bound state of fermions. So the remnant of the higher dimensional theory is not apparent. Nevertheless, there still remains some close relation between the LH and GHU scenarios as was mentioned in this subsection. (The correspondence may be rigorously argued once we utilize AdS-CFT correspondence.) 

There are various versions of the LH models. but here we discuss ``the simplest little Higgs" \cite{Schmaltz} in order to 
understand the key ingredient of the scenario, ``collective breaking". 

Interestingly, also in the LH the gauge symmetry should be enlarged. This is basically because the scenario needs a global symmetry which is larger than $SU(2) \times U(1)$, so that even after the Higgs mechanism physical NG bosons remain.  
Thus, let us consider the simplest SU(3) model with triplet scalar $\phi$. (The model is eventually extended to the one with additional U(1)$_X$. But for simplicity here we just ignore it.) The VEV 
\be 
\label{2.25}
\langle \phi \rangle = 
\begin{pmatrix}  
0 \\ 
0 \\ 
f
\end{pmatrix} 
\ee
causes a spontaneous breaking of global symmetry SU(3) \ $\to$ \ SU(2), and resultant NG bosons are written as 
\be 
\label{2.26} 
\pi = 
\begin{pmatrix} 
-\frac{\eta}{2} & 0 & \phi^{+} \\ 
0 & -\frac{\eta}{2} & \phi^{0} \\ 
\phi^{-} & \phi^{0 \ast} & \eta
\end{pmatrix},  
\ee 
where $h = (\phi^{+}, \ \phi^{0})^{t}$ is identified with the Higgs doublet. 
In this model, however, $h$ and $\eta$ are all absorbed to massive gauge bosons by the Higgs mechanism in the process of spontaneous breakdown SU(3) \ $\to$ \ SU(2), and no physical Higgs remains. 

Then we enlarge the model a little and introduce two copies of triplet scalars $\phi_{1}$ and $\phi_{2}$, and introduce covariant derivatives $D_{\mu}$ with the same SU(3) gauge bosons $A_{\mu}$ for both: 
\be
\label{2.27}  
{\cal L} = |D_{\mu} \phi_{1}|^{2} + |D_{\mu} \phi_{2}|^{2} \ \ \ 
( D_{\mu} = \partial_{\mu} -ig A_{\mu} ) 
\ee
where two triplet scalars are non-linearly realized: 
\be 
\label{2.28} 
\phi_{1} = e^{i\frac{\pi_{1}}{f}} 
\begin{pmatrix} 
0 \\  
f 
\end{pmatrix}, \ \ \ 
\phi_{2} = e^{i\frac{\pi_{2}}{f}} 
\begin{pmatrix} 
0 \\  
f 
\end{pmatrix}. 
\ee 
For simplicity the decay constants of $\phi_{1, 2}$ are taken to be the same: $f$. 

Each sector of two scalars $\phi_{1, 2}$ has its own global SU(3) symmetry. But, once gauge interaction is switched on, the global symmetry is explicitly broken to the ``diagonal" $SU(3)$; When each of 
$\phi_{1}$ and $\phi_{2}$ transforms as 
\be 
\label{2.29} 
\phi_{1} \ \to \ U_{1} \phi_{1}, \ \ \phi_2 \ \to \ U_{2} \phi_{2}
\ee 
gauge fields should transform as
\be 
\label{2.30}
A_{\mu} \ \to \ U_{1} A_{\mu} U_{1}^{\dagger}, \ \ \ 
A_{\mu} \ \to \ U_{2} A_{\mu} U_{2}^{\dagger}.  
\ee 
Since one gauge field $A_{\mu}$ couples with both triplets, for consistency it's necessary that $U_{1} = U_{2}$.

The NG boson corresponding to this diagonal SU(3) remains exactly massless, but is ``eaten" by the Higgs mechanism.  
The orthogonal one, in particular its doublet component $h$ is identified with Higgs and acquires a mass, since the relevant global symmetry is broken explicitly by the presence of the gauge interaction. 

Let us note that to break the global symmetry explicitly by the gauge interaction, SU(3)$^{2} \to$ SU(3), gauge couplings of $A_{\mu}$ to both of $U_{1}, \ U_{2}$ are necessary, Namely, the gauge couplings with both triplets collectively break the global symmetry. This is what ``collective breaking" means. The physically remaining Higgs is not real NG boson but a pseudo NG boson, acquiring a mass due to the breaking of the global symmetry. The collective breaking in turn means that even if Higgs acquires a mass at quantum level, it should be induced by Feynman diagrams where the gauge interactions with both triplets are included. Thus the diagram leading to the quadratic divergence of the Higgs mass is not allowed (at least at the one loop level) and Higgs does not suffer from the quadratically divergent quantum correction to the Higgs mass. This is the mechanism of collective breaking in order to remove the quadratic divergence. 

Let us note that in order for the collective breaking to work, the repetition of fields was essential, which seems to correspond to the presence of KK modes in GHU. Also, the quadratic divergence of Higgs mass disappears as the result of the repetition of fields in 
the LH model, just as in the case of GHU, where the sum over all KK modes provides a finite Higgs mass \cite{HIL}. 

\subsection{Close relation between GHU and Superstring} 

The GHU also has a very close relation to the superstring theory. In fact, the (bosonic part of) point particle limit of open superstring theory, i.e. 10D SUSY Yang-Mills theory, may be understood as a sort of GHU.  

Pure SUSY Yang-Mills theory with a gauge multiplet alone is possible only for specific space-time dimensionality $D = 3, 4, 6$ and 10, just because the matching of physical degrees of freedom of gauge boson and gauge fermion is realized in these dimensions: 
\be 
\label{2.31}  
D-2 = r \ 2^{[\frac{D-2}{2}]} \ \ (r = \frac{1}{2} \ \ \mbox{for Majorana-Weyl}). 
\ee 
This means that the 10D SUSY Yang-Mills theory (with $r = \frac{1}{2}$) is formulated as a pure SUSY Yang-Mills theory without any need of introducing additional matter field. Therefore, the only possible origin of the Higgs field is the gauge field, or to be more precise the extra-dimensional component of 10D gauge field. This is nothing but GHU.

Although it is usually argued that in supersymmetric theory the hierarchy problem of quadratic divergence is solved by the SUSY, if the theory is regarded as the one inspired by superstring, the hierarchy problem may be solved by the mechanism in the GHU as well. In fact, if the mode sum over all KK modes is performed, the quantum correction to the Higgs mass is finite even after the SUSY breaking.

\section{Normal vs. anomalous Higgs interactions} 

Now that a new particle, which behaves as the Higgs particle in the standard model (SM), has been discovered at CERN, only the theory with the SM as its decoupling limit should be acceptable as the theory of new physics. In order to determine which type of theory we should pursue among the remaining candidates, the precision test of the Higgs interaction is quite important. It may not be easy at LHC and we hope that it can be performed at planned ILC experiment. 

Interestingly, representative new physics theories show some sort of deviation of Higgs interactions from those predicted by the SM. 
For instance, in MSSM the couplings of the Higgs interactions generally depend on the parameters $\beta$ and $\alpha$. For instance, the ratios of the Yukawa coupling of the SM-like Higgs to that of the SM for the third generation are given as 
\bea 
&& \frac{f_{t}^{(MSSM)}}{f^{(SM)}_{t}} = \frac{\cos \alpha}{\sin \beta}, \\   
&& \frac{f_{b}^{(MSSM)}}{f^{(SM)}_{b}} = \frac{f_{\tau}^{(MSSM)}}{f^{(SM)}_{\tau}} = - \frac{\sin \alpha}{\cos \beta}. 
\label{3.1}
\eea 
Although the Yukawa couplings generally deviate from the predictions of the SM, the deviation is universal for all generations. It is an important prediction of the MSSM. It also should be noted that at the decoupling limit $\alpha \to \beta - \frac{\pi}{2}$, the deviations just go away, as we anticipate. 

Let us now discuss higher-dimensional theories. We will see that in the gauge-Higgs unification (GHU) and also in the related theories (dimensional deconstruction, little Higgs (LH) and probably superstring) we expect really ``anomalous" Higgs interactions, qualitatively different from those in the SM and MSSM, coming from the very fact that Higgs is not a scalar particle, but is originally a gauge field or a pseudo NG boson. 

From the viewpoint of the precision test of the Higgs interactions, it seems to be quite interesting that new physics theories are divided into two categories concerning the Higgs interactions, i.e. \\ 
・theories with normal Higgs interactions \\ 
・theories with anomalous Higgs interactions 

We discuss these two categories separately below.

\subsection{Theory with normal Higgs interactions} 

Higher dimensional new physics theory of this type we discuss here is the scenario of the universal extra dimension (UED) \cite{UED}. 
This theory is obtained by simply making the SM higher-dimensional, although to get a chiral theory the extra dimension is assumed to be 
orbifold, not just a circle or sphere. In particular, its Higgs sector is just as in the SM: no additional Higgs doublet is introduced in contrast to the case of MSSM and Higgs is just an elementary scalar field even in the bulk, in clear contrast to the case of GHU or LH. Thus the KK zero mode of the Higgs field has just the same Yukawa couplings as those in the SM. 

\subsection{Theories with anomalous Higgs interactions}

\noindent \underline{GHU} 

We first discuss gauge-Higgs unification (GHU), as a typical example of the new physics theory with anomalous Higgs interactions. 
In GHU (we work in the 5D space-time), SM-like Higgs $H$ is the zero-mode of the $A_{y}$, $A^{(0)}_{y}$, and has a physical meaning as the Wilson loop phase (AB phase), coming from the fact that the circle is a non-simply-connected space, as was discussed in the previous section: 
\be 
\label{3.2} 
W = e^{i \frac{g}{2} \oint A_{y} dy}  
= e^{ig_{4} \pi R A^{(0)}_{y}} 
 \ \ \ (\mbox{for the case of Abelian}), 
\ee 
where $R$ is the radius of the circle and in the line integral along the circle, only the zero mode contribution remains.  

The fact that Higgs should be regarded as a AB phase (or Wilson loop phase) as is seen in (\ref{3.2}) leads to the anomalous Higgs interactions. 

To see this, we start from a general argument on the fermion masses and Yukawa couplings that in any gauge theories with spontaneous gauge symmetry breaking, fermion mass term is written as 
\be 
\label{3.3}
m(v) \bar{\psi} \psi,  
\ee 
where $m(v)$ is a function, say ``mass function", of the VEV $v = \langle H \rangle$. 
The interaction of physical Higgs field $h$ with fermion is expected to be provided by a replacement 
$v \ \to \ v + h$, since the Higgs is the field to denote the shift of $H$ from its VEV. Thus, Yukawa coupling $f$ of $h$ with the (KK zero mode of) fermion is expected to be given by  
\be 
\label{3.4} 
f = \frac{d m(v)}{dv}.  
\ee 
This prescription perfectly works in the case of the SM, where $m(v)$ is a linear function of $v$, $m(v) = fv$. 
In the case of GHU, a story is a little more complicated. Even if we consider only one flavor, still fermion 
has infinite number of KK modes. Thus the mass term and Yukawa couplings are written in the form of matrices in the base of KK modes. Provided that the VEV $v$ is fixed, each KK mode should be given as an eigen-vector of the mass matrix determined by $v$. 
Thus in the base of KK modes the mass matrix (the matrix form of the mass terms) is diagonalized. Unusual thing, however, is that 
the matrix to denote the Yukawa couplings is generally non-diagonal even in the base of the mass eigen-states \cite{HKLT}. This characteristic feature of the GHU is closely related to the fact that the mass function $m(v)$ of KK zero mode, for instance, can be non-linear function such as trigonometric function, in general, reflecting that the Higgs is a phase, as we will see below. Thus, to be precise, the prescription 
given in (\ref{3.4}) is only for the diagonal elements of the Yukawa coupling matrix (valid for all KK modes, by the way).    

Since in the GHU the Higgs should be understood as a phase (AB phase), we expect that all physical observables have periodicity in the Higgs field $H$: 
\be 
\label{3.5}
v \ \ \to \ \ v + \frac{2}{g_{4}R} \ \ (g_{4}: \mbox{4D gauge coupling}). 
\ee 
In fact, we find for the zero mode quarks of light (1st and 2nd) generations (together with b quark), the quark masses are exponentially suppressed as in (\ref{2.7}), but in addition, their dependence on the VEV $v$ is approximately a trigonometric function:  
\be 
\label{3.6} 
m(v) \propto \sin (\frac{g_{4}}{2}\pi R v), 
\ee 
which is non-linear in $v$. Then, the prescription (\ref{3.4}) yields 
\be 
\label{3.7}  
f \propto \cos (\frac{g_{4}}{2}\pi R v). 
\ee 
It is remarkable fact that the Yukawa coupling even vanishes for a specific value of the VEV, 
\be 
\label{3.8} 
x \equiv \frac{g_{4}}{2}\pi R v = \frac{\pi}{2},  
\ee 
as was first claimed by Hosotani et al. \cite{HOOS}.

Although such a drastic case does not seem to be consistent with the recent data by LHC experiments, it is still possible that 
GHU predicts anomalous Yukawa couplings for light quarks for general value of $x$. In fact, the ratio of the Yukawa coupling predicted by the GHU to that predicted by the SM for light quarks is known to be very well approximated by an analytic formula \cite{HKLT}, 
\be 
\label{3.9}
\frac{f_{GHU}}{f_{SM}} \simeq x \cot x. 
\ee 
Note that at the ``decoupling limit" 　
\be 
\label{3.10} 
x = \frac{g_{4}}{2}v \pi R \ll 1 \leftrightarrow M_{W} \ll \frac{1}{R},  
\ee 
SM prediction is recovered, as is easily seen in (\ref{3.9}). 

Actually when the bulk mass $M_{i}$ is switched off, the mass function $m(v)$ becomes a linear function of $x$ as is seen in Fig.2(a), just as in the SM, so the Yukawa coupling does not deviate from the SM prediction for $x < \frac{\pi}{2}$. In this case, the periodicity of the mass eigenvalue is realized by a level crossing of the zero mode with the first KK mode at $x = \frac{\pi}{2}$ (see Fig.2(a)). Namely, at $x = \frac{\pi}{2}$ the zero mode is replaced by the first KK mode whose mass eigen-value decreases as $x$ grows up. For $M_{i} = 0$, however, there will not appear the mixing between these two modes because of the conservation of the (absolute value of) extra space component of the momentum. If we switch on the bulk mass $M_{i}$, the translational invariance along the extra space is violated by the presence of the bulk mass term (\ref{2.6}), thus causing a mixing between the two modes. Due to the mixing, the degeneracy of two mass eigen-values at the level crossing is lifted and there appears the deviation of the mass function from the linearity, leading to the trigonometric function for sufficiently large bulk mass $M_{i}$, as is seen in Fi.2(b). 

\begin{figure}[htb]
\begin{center}
\input{figure2.tex}\\ 
\end{center}
\end{figure}

Thus even though the original seed of the anomalous Higgs interaction is the periodicity characteristic to the GHU, more directly it may be attributed to the violation of translational invariance in the extra dimension. From this point of view, it may be worth noting that 
on the Randall-Sundrum background, the translational invariance is ``always" (universally) violated by the presence of the warp factor 
$e^{-\kappa |y|}$. This should be the reason why on the Randall-Sundrum background the Higgs interactions with massive gauge bosons, $W$ and $Z$, are also anomalous \cite{HS}, while on the flat space-time it is ``almost normal" (normal for $x < \frac{\pi}{2}$), as in the gauge-Higgs sector there is no parameter violating translational invariance in the case of flat space-time \cite{HKLT}. 

From field theoretical point of view, the reason why we get such non-linearity in the mass function like (\ref{3.6}) is that, due to the non-diagonal Yukawa couplings mentioned earlier, there appear mixings between zero mode and non-zero modes, which leads to the higher mass dimensional operators at the tree level of the form $h^{2n+1}\bar{\psi}\psi \ (n = 1,2, \ldots)$ (just like the dimension 5 operator in the see-saw mechanism). The coefficient of those higher mass-dimensional operators are suppressed by the inverse powers of $M_{c} = \frac{1}{R}$ and 
as we have already seen the effects of such ``irrelevant" operators go away at the decoupling limit $M_{c} \to \infty$. 

\subsection{Dimensional deconstruction} 

As we have seen in the previous section, though it's formulated on the ordinary 4D space-time, the scenario of the dimensional deconstruction may be understood as a ``latticized" GHU. We thus expect that also in this scenario, the anomalous Higgs interaction 
is expected. To be more specific, in this case, even if we adopt just periodic boundary conditions for fields like $A^{N+1}_{\mu} = A^{1}_{\mu}$, corresponding to the choice of a circle, not an orbifold, as the extra dimension in GHU, still the latticizing itself breaks the (continuous) translational invariance, which should lead to anomalous Higgs interactions. In fact, the mass eigen-values in this scenario generally behave as trigonometric functions: 
\be 
\label{3.11}
m_{n}(v) = \frac{2}{a} \sin( \frac{n\pi}{N} + \frac{g_{4}a v}{4}) \ \ \ (a: \mbox{lattice spacing}) 
\ee 
which are basically the same as the eigen-frequencies of the system of springs and balls. As you can easily see (\ref{3.11}) reduces to linear functions of $v$ in the continuum limit $a \to 0, \ N \to \infty$, keeping the circumference of the circle $L \equiv Na$ a constant; $m_{n}(v) \to \frac{2\pi}{L}n + \frac{g_{4}v}{2}$. A work on the anomalous interactions in the dimensional deconstruction along this line of argument is now in progress \cite{KLT}.         

\subsection{Little Higgs} 

We also have argued that the scenario of little Higgs (LH) has a close relationship with GHU. 
Just as in the case of GHU, the Higgs is nonlinearly realized, 
\be 
\label{3.12} 
U =e^{i\frac{H}{v}}, 
\ee
and therefore periodicity and non-linearity concerning the Higgs filed seem to arise very naturally. Thus we again expect anomalous Higgs interactions in the scenario of LH, though more complete analysis are clearly desirable in this case.

\section{$H \to \gamma \gamma$} 

$H \to \gamma \gamma$ is the clean decay mode of the Higgs, important for the identification of the Higgs particle. This di-photonic decay is induced at the quantum level since the photon is a massless particle, and is a very good testing ground of new physics. It also is sensitive to the Higgs interactions with fermions and gauge bosons running inside the loop diagrams, thus providing useful information on the property of Higgs interactions, whether they are normal or anomalous, as was discussed in some details in the previous section. The gluon fusion process $gg \to H$, a main process of the Higgs production, is also induced at the quantum level and has some similarity to the photonic Higgs decay, though gauge bosons do not contribute in this case. 

In this section we focus on two higher dimensional new physics scenarios having considerably different properties as we have already seen 
in the previous sections: UED and GHU. We will pay much attention on the difference of predictions these two scenarios make on the photonic decay.    

Recently, there has been an interesting claim that the contribution of the non-zero KK modes of top quark to the decay amplitude of $H \to \gamma \gamma$ has an opposite sign to that of the top quark in the SM and to that of the KK modes of top quark in UED \cite{MO}. The origin of such qualitative difference may be rather easily understood once we rely on the operator analysis. Let's assume $M_{c} = 1/R \gg v$. At the first glance the relevant operator for the photonic decay seems to be $h F_{\mu \nu}F^{\mu \nu}$ with $F_{\mu \nu}$ being the field strength of the photon. If this is the case, the coefficient should be divergent in the process of KK mode sum, since the mass dimension of the operator is 5. 
But this operator is not gauge invariant under SU(2)$_{L}\times$ U(1)$_{Y}$. So, actually the decay amplitude is dominated by 
a gauge invariant operator of mass dimension 6 written in terms of the Higgs doublet $\phi$: 
\be 
\label{4.1} 
\phi^{\dagger}\phi F_{\mu \nu}F^{\mu \nu}.  
\ee 
Note that when one of the Higgs doublets is replaced by its VEV, this yields the dimension 5 operator mentioned above. Now the coefficient of the operator (\ref{4.1}) is finite at least for 5D space-time. 

Also note that the operator in (\ref{4.1}) comes from the diagram shown in Fig.3, obtained by inserting Higgs doublet $\phi$ twice to the self-energy diagram of photon, quantum correction to the operator $F_{\mu \nu}F^{\mu \nu}$. So by utilizing the background field method, 
the Wilson coefficient of the operator (\ref{4.1}) is obtainable just calculating the self-energy diagram with the top quark (or its KK modes) inside the loop having the masses written by the VEV $v$, and finally taking the second derivative with 
respect to $v$. Since the coefficient of the self-energy diagram depends on the mass of the virtual state logarithmically, 
the coefficient of (\ref{4.1}) is given as 
\be 
\label{4.2} 
C \propto \frac{d^{2}}{dv^{2}} \log m_{n}^{2}|_{v = 0},  
\ee
where $m_{n}$ is the mass eigen-values of the $n$-th KK mode. 

\begin{figure}[tb]
\centering 
\includegraphics[width=8cm]{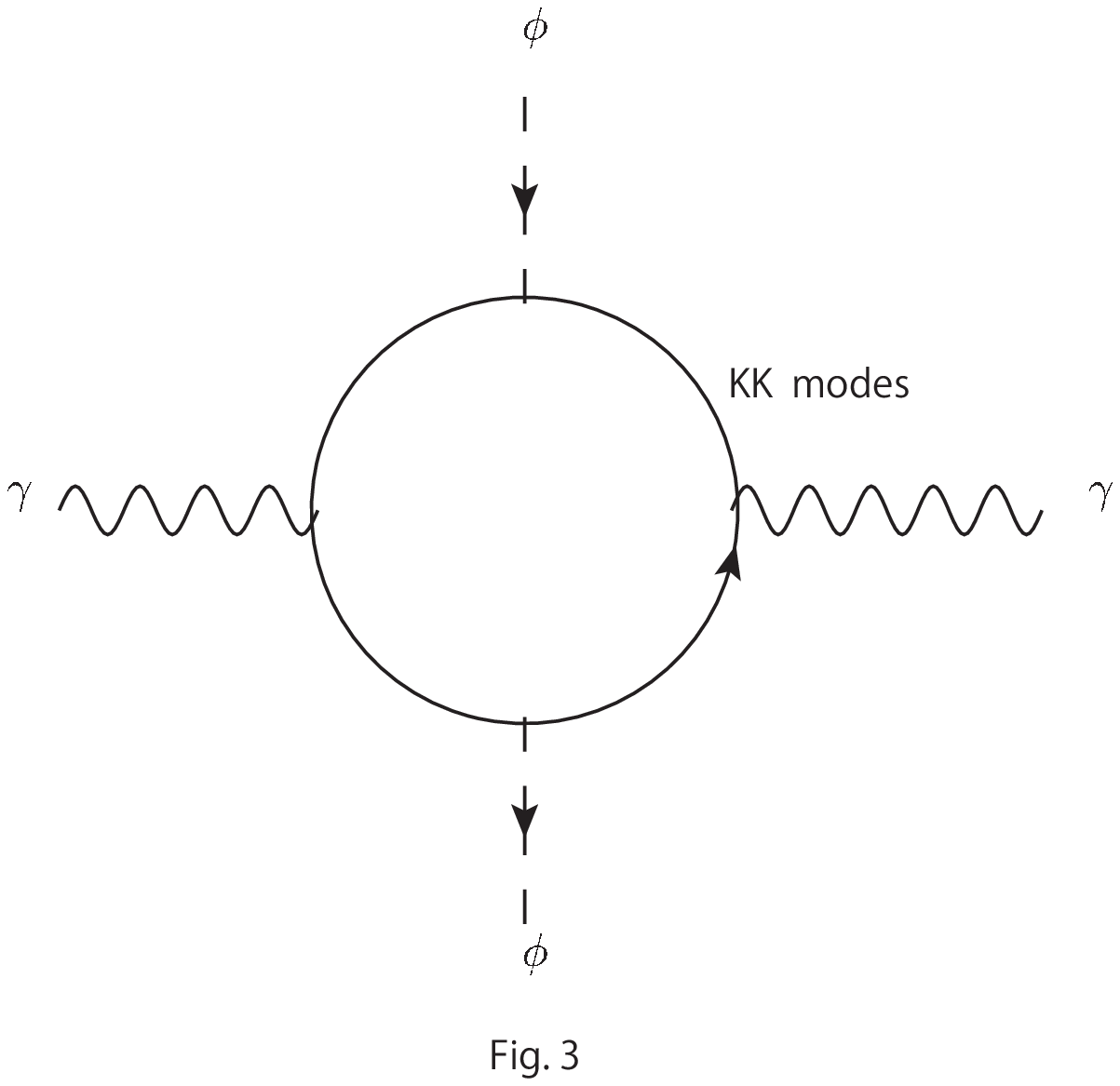}  
\end{figure}%

Now we are ready to discuss the difference between the predictions of UED and GHU. In each scenario, the mass eigen-value of a fermion is given as follows: 
\be 
\label{4.3} 
m_{n} = 
\begin{cases} 
\sqrt{(\frac{n}{R})^{2} + (fv)^{2}} \ \ \mbox{for} \ \ \mbox{UED} \\ 
\frac{n}{R} + fv \ \ \mbox{for} \ \ \mbox{GHU}      
\end{cases}. 
\ee
Here for brevity, we have ignored the possible bulk mass and $f$ is the Yukawa coupling for zero mode fermion. Then the procedure of (\ref{4.2}) gives just opposite results in sign for two scenarios: 
\be 
\label{4.4} 
C \propto 
\begin{cases}  
2\frac{f^{2}}{(\frac{n}{R})^{2}} \ \ \mbox{for} \ \ \mbox{UED} \\ 
-2 \frac{f^{2}}{(\frac{n}{R})^{2}} \ \ \mbox{for} \ \ \mbox{GHU}   
\end{cases}. 
\ee 

There is a claim that the new physics contributions of the theories which have the mechanism to solve the hierarchy problem invoking some symmetry have a tendency that the amplitude of gluon fusion process, for instance, is reduced. Intuitively, the reason might be the following. In the case of gluon fusion, the field strength in (\ref{4.1}) is of the gluon field. Since QCD sector has no direct connection with the Higgs sector, even if we take off the gluonic field the operator is still gauge invariant and it just becomes the mass-squared operator for the Higgs field. On the other hand any theories which aim to solve the problem of quadratic divergence of the Higgs mass should have a mechanism to reduce the quantum correction to the Higgs mass-squared by introducing, e.g. super-partners, KK modes, etc. 
Thus it will be natural to expect some destructive contributions from new heavy particles in this type of theories. 

Let us also note that the operator (\ref{4.1}) is easily generalized by replacing $\phi^{\dagger}\phi$ by its arbitrary function $V(\phi)$: i.e. $V(\phi)F^{\mu \nu}F_{\mu \nu}$. If we take off $F^{\mu \nu}F_{\mu \nu}$ again, we just get a Higgs potential $V(\phi)$. On the other hand, in the GHU the potential is completely finite \cite{Hosotani}, \cite{HIL}. Maru thus claims that the decay amplitude of $H \to \gamma \gamma$ is finite irrespectively of the space-time dimension \cite{Maru}.  

For more recent detailed analysis concerning the di-photonic decay in GHU, refer to \cite{MO2}. 
For a recent argument on the contribution of the non-zero KK modes of top quark in the SO(5) x U(1) GHU on the Randall-Sundrum 
background, refer to \cite{FHHOS}. 

Usually, the contributions of lighter generations (1st and 2nd generations) to the photonic Higgs decay is more or less negligible, since their Yukawa couplings are strongly suppressed by very small quark masses. That is why only the contribution of the top quark is seriously taken into account. In the GHU, however, the story is not so simple. Namely, the contribution of each non-zero KK mode of lighter generation of quark is not strongly suppressed by the corresponding small quark masses. This is because only the mode function of the zero mode exhibits the localization at a fixed point and the Yukawa couplings of non-zero KK modes are not exponentially suppressed, being of the order of gauge coupling. So we naturally expect possible significant contributions to the photonic decay from lighter generations as well.   
   
Fortunately or unfortunately, it turns out that although each KK mode gives significant contribution to the decay amplitude, as we expected, after the summation over the contributions of all KK modes, the factor suppressed by $1/M_{c}^{2} = 1/(1/R)^{2}$ (what we expect from decoupling theorem) just goes away, and remaining contribution is exponentially suppressed as $e^{- M_{i}R} \ (M_{i}: \mbox{bulk mass})$ \cite{HKL}. 

We recall that a similar exponential suppression factor due to relatively large bulk mass was found in the process of the calculation of the quantum correction to the Higgs mass-squared \cite{HIL}. It may not be so surprising, since as we have argued above, the operator relevant for the photonic decay has some similarity to that ot the Higgs potential. We may also notice that the exponential factor mimics the Boltzman factor in the case of finite temperature field theory. This factor goes away at de-compactification limit $R \to \infty$, thus suggesting that it comes from non-local gauge invariant operator due to the Wilson loop. In fact, this should be the reason why we get the finite decay amplitude in GHU irrespectively of the dimensionality of the space-time as was demonstrated in the paper \cite{Maru}. 
We may also understand that the factor is the contribution of the sector with non-zero winding numbers at the Poisson re-summation which clearly shows the contribution of the distances larger than $R$ and thus suppressed by the exponential factor $e^{-M_{i}R}$, present in the Yukawa potential.  
 
\section{Summary} 

In spite of the great success at LHC experiments to have discovered the particle consistent with the Higgs of the SM, it is still premature to conclude that the particle is the Higgs of the SM. It may be some SM-like Higgs particle some new physics theory predicts. Also the Higgs sector of the SM is still mysterious having a few important long-standing problems, such as hierarchy problem and the problem of too many arbitrary parameters present in the sector. Thus we should continue to ask what is the origin of the Higgs. 

In this article we focused on higher dimensional new physics theories. Also discussed was 4-dimensional theories, such as little Higgs (LH) and dimensional deconstruction, which possess some close relationship with the higher dimensional theory, especially with the gauge-Higgs unification (GHU). Important common features shared by these scenarios are the presence of the ``shift symmetry", which guarantees the finiteness of the Higgs mass at the quantum level, and the periodicity: physical observables are periodic in the Higgs field. In GHU the origin of the Higgs field is a gauge field and in the LH and dimensional deconstruction the origin is a (pseudo) Nambu-Goldstone boson.  

In order to investigate the origin of the Higgs, the precision tests of Higgs interactions will be of crucial importance. We have discussed that new physics theories are divided into two categories: theories with normal Higgs interactions, and theories with 
anomalous Higgs interactions. 

From this point of view, we have demonstrated that in the scenario of GHU and related LH and dimensional deconstruction scenarios, the periodicity mentioned above (together with the breakdown of translational invariance along the extra dimension) leads to very characteristic anomalous Higgs interactions, which are never shared by the SM. That should be very important observation in the sense to understand deeply the origin of the Higgs particle. 

The difference, i.e. normal (as in the case of UED) or anomalous Higgs interactions, also makes the new physics contribution to the main decay mode of the Higgs, $H \to \gamma \gamma$, qualitatively different. 

Such crucial precision tests of the Higgs interactions may not be easy at LHC experiment, but we hope that it will be performed at the planned ILC experiment.

\subsection*{Acknowledgments}

The author would like to thank Y. Hosotani, N. Maru and Y. Sakamura for very useful
discussions and the explanation of their works, which are relevant to this article. 
This work was supported in part by the Grant-in-Aid for Scientific Research of the Ministry of Education, Science and Culture, No.~21244036, No.~23654090, No.~23104009.

\end{document}

%% file: figure1.tex
{\unitlength 0.1in
\begin{picture}( 40.4000, 15.9000)( 19.6000,-23.9000)
%
\special{pn 8}%
\special{ar 3050 930 130 130  1.5707963  1.4056476}%
%
\special{pn 8}%
\special{ar 4020 930 122 122  0.0000000  6.2831853}%
\put(39.3000,-9.9000){\makebox(0,0)[lb]{$G_{s}$}}%
\put(29.6000,-9.9000){\makebox(0,0)[lb]{$G$}}%
%
\special{pn 8}%
\special{ar 4980 930 120 120  0.0000000  6.2831853}%
\put(49.1000,-9.9000){\makebox(0,0)[lb]{$G$}}%
%
\special{pn 8}%
\special{pa 3200 940}%
\special{pa 3890 940}%
\special{fp}%
%
\special{pn 8}%
\special{pa 4150 940}%
\special{pa 4850 940}%
\special{fp}%
%
\special{pn 8}%
\special{pa 2940 980}%
\special{pa 2410 1330}%
\special{fp}%
%
\special{pn 8}%
\special{ar 2330 1410 132 132  0.0000000  6.2831853}%
%
\special{pn 8}%
\special{pa 5100 960}%
\special{pa 5560 1320}%
\special{fp}%
%
\special{pn 8}%
\special{ar 5640 1420 120 120  0.0000000  6.2831853}%
%
\special{pn 8}%
\special{pa 2240 1490}%
\special{pa 1960 2010}%
\special{dt 0.045}%
%
\special{pn 8}%
\special{pa 5730 1480}%
\special{pa 6000 1990}%
\special{dt 0.045}%
%
\special{pn 8}%
\special{pa 2600 1210}%
\special{pa 2760 1100}%
\special{dt 0.045}%
\special{sh 1}%
\special{pa 2760 1100}%
\special{pa 2694 1122}%
\special{pa 2716 1130}%
\special{pa 2716 1154}%
\special{pa 2760 1100}%
\special{fp}%
%
\special{pn 8}%
\special{pa 3460 940}%
\special{pa 3660 940}%
\special{dt 0.045}%
\special{sh 1}%
\special{pa 3660 940}%
\special{pa 3594 920}%
\special{pa 3608 940}%
\special{pa 3594 960}%
\special{pa 3660 940}%
\special{fp}%
%
\special{pn 8}%
\special{pa 4400 950}%
\special{pa 4680 930}%
\special{dt 0.045}%
\special{sh 1}%
\special{pa 4680 930}%
\special{pa 4612 916}%
\special{pa 4628 934}%
\special{pa 4616 956}%
\special{pa 4680 930}%
\special{fp}%
%
\special{pn 8}%
\special{pa 5250 1090}%
\special{pa 5430 1220}%
\special{dt 0.045}%
\special{sh 1}%
\special{pa 5430 1220}%
\special{pa 5388 1166}%
\special{pa 5388 1190}%
\special{pa 5364 1198}%
\special{pa 5430 1220}%
\special{fp}%
\put(22.2000,-14.8000){\makebox(0,0)[lb]{$G_{s}$}}%
\put(55.5000,-14.7000){\makebox(0,0)[lb]{$G_{s}$}}%
\put(39.3000,-25.2000){\makebox(0,0)[lb]{Fig. 1}}%
\end{picture}}%

%% file: figure2.tex
{\unitlength 0.1in
\begin{picture}( 42.1000, 33.0000)( 13.7000,-81.4000)
%
\special{pn 8}%
\special{pa 3800 7400}%
\special{pa 5400 7400}%
\special{fp}%
\special{sh 1}%
\special{pa 5400 7400}%
\special{pa 5334 7380}%
\special{pa 5348 7400}%
\special{pa 5334 7420}%
\special{pa 5400 7400}%
\special{fp}%
%
\special{pn 8}%
\special{pa 3800 7400}%
\special{pa 3800 5200}%
\special{fp}%
\special{sh 1}%
\special{pa 3800 5200}%
\special{pa 3780 5268}%
\special{pa 3800 5254}%
\special{pa 3820 5268}%
\special{pa 3800 5200}%
\special{fp}%
%
\special{pn 8}%
\special{pa 3810 7400}%
\special{pa 3960 7340}%
\special{pa 3990 7330}%
\special{pa 4020 7322}%
\special{pa 4050 7312}%
\special{pa 4082 7304}%
\special{pa 4144 7286}%
\special{pa 4174 7278}%
\special{pa 4204 7268}%
\special{pa 4236 7260}%
\special{pa 4266 7252}%
\special{pa 4298 7246}%
\special{pa 4330 7242}%
\special{pa 4426 7232}%
\special{pa 4490 7230}%
\special{pa 4522 7230}%
\special{pa 4618 7236}%
\special{pa 4648 7242}%
\special{pa 4712 7254}%
\special{pa 4742 7262}%
\special{pa 4774 7268}%
\special{pa 4804 7278}%
\special{pa 4866 7296}%
\special{pa 4898 7306}%
\special{pa 4928 7316}%
\special{pa 4986 7336}%
\special{pa 5012 7344}%
\special{pa 5044 7356}%
\special{pa 5088 7370}%
\special{pa 5126 7382}%
\special{pa 5124 7376}%
\special{pa 5110 7370}%
\special{fp}%
%
\special{pn 8}%
\special{pa 3800 6800}%
\special{pa 3828 6816}%
\special{pa 3858 6830}%
\special{pa 3916 6856}%
\special{pa 3946 6868}%
\special{pa 3976 6878}%
\special{pa 4006 6886}%
\special{pa 4162 6926}%
\special{pa 4194 6934}%
\special{pa 4224 6938}%
\special{pa 4256 6944}%
\special{pa 4288 6946}%
\special{pa 4320 6952}%
\special{pa 4384 6960}%
\special{pa 4414 6966}%
\special{pa 4478 6970}%
\special{pa 4510 6970}%
\special{pa 4574 6966}%
\special{pa 4606 6962}%
\special{pa 4670 6950}%
\special{pa 4700 6944}%
\special{pa 4764 6928}%
\special{pa 4824 6912}%
\special{pa 4854 6902}%
\special{pa 4886 6892}%
\special{pa 4916 6882}%
\special{pa 4978 6864}%
\special{pa 5008 6852}%
\special{pa 5036 6838}%
\special{pa 5064 6822}%
\special{pa 5090 6806}%
\special{fp}%
%
\special{pn 8}%
\special{pa 3806 6806}%
\special{pa 3830 6784}%
\special{pa 3904 6724}%
\special{pa 3932 6706}%
\special{pa 3958 6690}%
\special{pa 3986 6674}%
\special{pa 4016 6660}%
\special{pa 4046 6648}%
\special{pa 4076 6638}%
\special{pa 4138 6620}%
\special{pa 4168 6608}%
\special{pa 4196 6596}%
\special{pa 4226 6586}%
\special{pa 4290 6574}%
\special{pa 4320 6566}%
\special{pa 4352 6558}%
\special{pa 4382 6550}%
\special{pa 4414 6548}%
\special{pa 4542 6556}%
\special{pa 4606 6562}%
\special{pa 4638 6566}%
\special{pa 4670 6572}%
\special{pa 4762 6596}%
\special{pa 4794 6606}%
\special{pa 4824 6616}%
\special{pa 4854 6628}%
\special{pa 4882 6642}%
\special{pa 4966 6690}%
\special{pa 4992 6706}%
\special{pa 5018 6726}%
\special{pa 5042 6748}%
\special{pa 5066 6768}%
\special{pa 5090 6786}%
\special{fp}%
%
\special{pn 8}%
\special{pa 3800 6200}%
\special{pa 3820 6226}%
\special{pa 3840 6250}%
\special{pa 3864 6272}%
\special{pa 3948 6320}%
\special{pa 3976 6334}%
\special{pa 4036 6358}%
\special{pa 4066 6366}%
\special{pa 4098 6374}%
\special{pa 4130 6380}%
\special{pa 4160 6386}%
\special{pa 4192 6392}%
\special{pa 4224 6400}%
\special{pa 4254 6406}%
\special{pa 4350 6424}%
\special{pa 4380 6428}%
\special{pa 4444 6436}%
\special{pa 4476 6438}%
\special{pa 4540 6434}%
\special{pa 4604 6426}%
\special{pa 4634 6420}%
\special{pa 4762 6396}%
\special{pa 4792 6388}%
\special{pa 4822 6378}%
\special{pa 4880 6352}%
\special{pa 4910 6336}%
\special{pa 4994 6294}%
\special{pa 5024 6280}%
\special{pa 5108 6232}%
\special{pa 5120 6226}%
\special{fp}%
%
\special{pn 8}%
\special{pa 3796 6210}%
\special{pa 3822 6194}%
\special{pa 3990 6098}%
\special{pa 4046 6070}%
\special{pa 4106 6042}%
\special{pa 4134 6030}%
\special{pa 4194 6008}%
\special{pa 4226 5998}%
\special{pa 4288 5982}%
\special{pa 4318 5976}%
\special{pa 4350 5970}%
\special{pa 4382 5962}%
\special{pa 4412 5954}%
\special{pa 4444 5948}%
\special{pa 4476 5946}%
\special{pa 4508 5946}%
\special{pa 4540 5952}%
\special{pa 4570 5958}%
\special{pa 4602 5966}%
\special{pa 4632 5974}%
\special{pa 4664 5984}%
\special{pa 4694 5996}%
\special{pa 4752 6020}%
\special{pa 4810 6048}%
\special{pa 4838 6062}%
\special{pa 4866 6078}%
\special{pa 4892 6094}%
\special{pa 4920 6112}%
\special{pa 4974 6146}%
\special{pa 5056 6200}%
\special{pa 5082 6218}%
\special{pa 5100 6230}%
\special{fp}%
%
\special{pn 8}%
\special{pa 3800 5606}%
\special{pa 3826 5624}%
\special{pa 3878 5664}%
\special{pa 3906 5682}%
\special{pa 3958 5718}%
\special{pa 3986 5736}%
\special{pa 4042 5768}%
\special{pa 4098 5796}%
\special{pa 4128 5808}%
\special{pa 4188 5828}%
\special{pa 4280 5852}%
\special{pa 4312 5860}%
\special{pa 4376 5872}%
\special{pa 4406 5876}%
\special{pa 4470 5880}%
\special{pa 4502 5880}%
\special{pa 4566 5876}%
\special{pa 4630 5864}%
\special{pa 4662 5856}%
\special{pa 4722 5840}%
\special{pa 4754 5828}%
\special{pa 4784 5818}%
\special{pa 4812 5806}%
\special{pa 4842 5794}%
\special{pa 4872 5780}%
\special{pa 4928 5752}%
\special{pa 4958 5738}%
\special{pa 5014 5706}%
\special{pa 5042 5692}%
\special{pa 5126 5644}%
\special{pa 5130 5640}%
\special{fp}%
\put(55.8000,-74.0000){\makebox(0,0)[lb]{$x$}}%
\put(37.9500,-49.7000){\makebox(0,0)[lb]{$m_{n}$}}%
%
\special{pn 8}%
\special{pa 1400 7400}%
\special{pa 3000 7400}%
\special{fp}%
\special{sh 1}%
\special{pa 3000 7400}%
\special{pa 2934 7380}%
\special{pa 2948 7400}%
\special{pa 2934 7420}%
\special{pa 3000 7400}%
\special{fp}%
%
\special{pn 8}%
\special{pa 1400 7400}%
\special{pa 1400 7400}%
\special{fp}%
\special{pa 1400 7400}%
\special{pa 1400 5200}%
\special{fp}%
\special{sh 1}%
\special{pa 1400 5200}%
\special{pa 1380 5268}%
\special{pa 1400 5254}%
\special{pa 1420 5268}%
\special{pa 1400 5200}%
\special{fp}%
%
\special{pn 8}%
\special{pa 1400 7400}%
\special{pa 2820 6820}%
\special{fp}%
%
\special{pn 8}%
\special{pa 1400 6800}%
\special{pa 2800 7400}%
\special{fp}%
%
\special{pn 8}%
\special{pa 1400 6810}%
\special{pa 2800 6200}%
\special{fp}%
\put(13.7000,-50.0000){\makebox(0,0)[lb]{$m_{n}$}}%
\put(32.2000,-74.0000){\makebox(0,0)[lb]{$x$}}%
\put(21.2000,-76.9000){\makebox(0,0)[lb]{$\frac{\pi}{2}$}}%
\put(45.1000,-76.7000){\makebox(0,0)[lb]{$\frac{\pi}{2}$}}%
%
\special{pn 8}%
\special{pa 1400 6200}%
\special{pa 2800 6800}%
\special{fp}%
%
\special{pn 8}%
\special{pa 1400 6200}%
\special{pa 2800 5600}%
\special{fp}%
%
\special{pn 8}%
\special{pa 1400 5600}%
\special{pa 2800 6200}%
\special{fp}%
\put(21.2000,-79.6000){\makebox(0,0)[lb]{(a)}}%
\put(45.1000,-79.7000){\makebox(0,0)[lb]{(b)}}%
\put(33.4000,-82.7000){\makebox(0,0)[lb]{Fig. 2}}%
%
\special{pn 8}%
\special{pa 5120 7390}%
\special{pa 5120 5260}%
\special{dt 0.045}%
%
\special{pn 8}%
\special{pa 2780 7400}%
\special{pa 2780 5240}%
\special{dt 0.045}%
\end{picture}}%